\def\IsMainDocument{}

\documentclass[%
a4paper,
aps,
prl,
twocolumn,
nofootinbib,
amsmath,amssymb,
superscriptaddress,
10pt,
floatfix,
]{revtex4-2}

\usepackage[a4paper,centering,hmargin=2.25cm,vmargin=1.75cm]{geometry}

\usepackage{graphicx}
\usepackage{amsfonts,amsmath}
\usepackage{mathtools}
\usepackage{amsmath}
\usepackage{xcolor}
\usepackage{bbold}
\usepackage{xr} 
\usepackage[colorlinks=true,citecolor=blue,linkcolor=blue,urlcolor=blue,filecolor=black]{hyperref}
\usepackage{physics}
\usepackage{braket}
\usepackage{ulem}
\usepackage{comment}
\usepackage{afterpage}
\usepackage{cancel}

\usepackage{microtype}

\usepackage[capitalize]{cleveref}

\let\emph\relax
\DeclareTextFontCommand{\emph}{\itshape}

\def\bbI{\mathbb{I}}
\def\bbO{\mathcal{O}}
\def\bbQ{\mathcal{Q}}
\def\TM{\mathcal{T}}
\def\tmE{\mathbb{E}^T}
\def\tmEp{\mathbb{E}^{T+1}}

\def\rank{\mathcal{R}}
\def\Lt{\Lambda^{s}}  
\def\Lb{\Lambda^{o}}
\def\Lbt{{\bar\Lambda}^{s}}  
\def\Lbb{{\bar\Lambda}^{o}}
\def\Ts{{\mathcal T}}

\newcommand{\Eq}[1]{{Eq.~({\ref{#1}})}}
\newcommand{\Fig}[1]{{Fig.~{\ref{#1}}}}

\newcommand{\beq}{\begin{equation}}
\newcommand{\eeq}{\end{equation}}

\ifdefined\IsMainDocument
    \newcommand{\smref}[1]{\ref{#1}}
\else
    \newcommand{\smref}[1]{\textcolor{black}{\ref*{#1}}}
\fi

\begin{document}

\title{Mesoscopic Regimes of Temporal Entanglement in Ergodic Quantum Systems}

\author{Sergio Cerezo-Roquebrún}
\affiliation{Instituto de Física Teórica UAM/CSIC, C/ Nicolás Cabrera 13--15, Cantoblanco, 28049 Madrid, Spain}
\author{Jan Thorben Schneider}
\affiliation{Institute of Fundamental Physics IFF-CSIC, C/ Serrano 113b, Madrid 28006, Spain}%
\author{Stefano Carignano}
\affiliation{Barcelona Supercomputing Center, 08034 Barcelona, Spain}
\author{Aleix Bou-Comas}
\affiliation{Institute of Fundamental Physics IFF-CSIC, C/ Serrano 113b, Madrid 28006, Spain}%
\author{Mari Carmen Bañuls}
\affiliation{Max-Planck-Institut fur Quantenoptik, Hans-Kopfermann-Str.~1, D-85748 Garching, Germany}
\affiliation{Munich Center for Quantum Science and Technology (MCQST), Schellingstr.~4, D-80799 München, Germany}
\author{Esperanza López}
\affiliation{Instituto de Física Teórica UAM/CSIC, C/ Nicolás Cabrera 13--15, Cantoblanco, 28049 Madrid, Spain}%

\author{Luca Tagliacozzo}
\affiliation{Institute of Fundamental Physics IFF-CSIC, C/ Serrano 113b, Madrid 28006, Spain}
\email{luca.tagliacozzo@iff.csic.es}

\date{\today}

\begin{abstract}
We study temporal correlations in interacting quantum systems through the influence functional of a half-infinite quantum Ising chain. Using Rényi entropies and temporal mutual information, we confirm that integrable dynamics is captured by the quasiparticle picture. In contrast, generic ergodic Hamiltonian dynamics exhibits pronounced deviations from random-circuit universality, and its generalization including a symmetry accounting for energy conservation.
Instead, we find a long mesoscopic regime 
suggestive of a slow spectral reorganization of the influence functional. Our results reveal a rich temporal structure in generic Hamiltonian dynamics and point to limitations of existing random-circuit paradigms at experimentally and numerically relevant timescales.

\end{abstract}

\maketitle


\emph{Introduction. --- } The real-time dynamics of interacting quantum systems remains one of the central challenges in many-body physics. 
While local observables often relax to thermal values, a microscopic understanding of how quantum correlations spread and reorganize in time is still incomplete.

Recent progress has been driven by exactly solvable models of local ergodic dynamics, including dual-unitary and random quantum circuits~\cite{bertini2018,bertini2019,gopalakrishnan2019,claeys2020,fisher2023}. 
These systems exhibit a universal pattern of entanglement growth described by an effective membrane picture~\cite{nahum2017,nahum2018,jonay2018coarse,zhou2019emergent,zhou2020membrane}, a mechanism fundamentally distinct from the quasiparticle spreading governing integrable and conformal field theories~\cite{calabrese_2005,alba2017quench}. 
To model ergodic Hamiltonian dynamics, where energy conservation is fundamental, random circuits can be equipped with a local conserved quantity,
which gives rise to a slower diffusive growth of all $n$-Rényi entropies with $n>1$
\cite{rakovszky2018,rakovszky2019,znidaric2020,zhou2020diffusive,rakovszky2021,huang2020a,huang2022}. 
This framework has successfully accounted for a wide range of dynamical phenomena and is often considered a minimal model for the behavior of generic interacting Hamiltonian dynamics at late times.

Space-time duality facilitates a different perspective, especially useful for one-dimensional systems.
In this case, essential dynamical information can be encoded in a temporal operator representing the influence functional of part of the system, giving rise to the idea of temporal entanglement~\cite{banuls2009,sonner2021aop,cerezo2025,keeling2025pt,bertini2025exactlysolvablemanybodydynamics}.
The duality has been used to connect the scaling of entanglement in the time and space directions in circuit settings~\cite{lu2021,ippoliti2022}. Different notions of temporal and generalized temporal entanglement have recently emerged in diverse areas from quantum information  to many-body quantum systems, field theories and holography ~\cite{leifer2013,narayan2015,pollock2018,cotler2018,sonner2021aop,carignano2024,carignano2025a,narayan2023a,nakata2021,doi2023a,li2023,heller2025a,heller2025b,nunez2025a,nunez2025b,bou2024,liu2025,takayanagi2025,castro-alvaredo2026}.
Furthermore, the universal scaling of temporal correlations has been understood for dual unitaries and random circuits~\cite{foligno2023}, whereas for Hamiltonian dynamics different regimes have been numerically observed for integrable and non-integrable cases~\cite{muller-hermes2012,lerose2021,Giudice_PhysRevLett.128.220401,carignano2024}, but a generic understanding is missing.

A key open question is therefore whether the random-circuit paradigm, possibly enriched by a conserved quantity, can quantitatively capture the temporal correlations generated by generic Hamiltonian dynamics.
We address this question in a one-dimensional quantum Ising chain with transverse and longitudinal fields, which interpolates between integrable and strongly non-integrable regimes~\cite{kim2013}. We characterize the spreading of temporal correlations through Rényi entropies and mutual information of the influence functional associated with a half-infinite system, which we represent efficiently as a temporal matrix product state~\cite{banuls2009, muller-hermes2012,hastings2015,tirrito2018characterizing,frias2022,lerose2023overcoming,carignano2024a}. This approach allows us to access infinite-system dynamics at finite, but moderately long times.

We find that integrable dynamics
agrees with the quasiparticle picture, with linear growth of Rényi entropies and saturation of temporal mutual information.~\cite{calabrese_2005}. 
In contrast, non-integrable cases show significant deviations from random-circuit models across the entire accessible time window, featuring a long mesoscopic timescale during which temporal correlations develop a complex structure.
In particular, Rényi entropies display sublinear growth
and 
ratios of moments of the influence functional
exhibit non-monotonic behavior. At a qualitative level, this can be understood from a simple toy-model  for the spectrum of the influence functional: a broad background of many small eigenvalues coexists with a few leading eigenvalues that progressively separate from the bulk.
However, a simple diffusive or power law leading eigenvalue are not enough to explain our observations.
The long metastable regime we observe could thus reflect a slow reorganization of the spectrum, rather than the immediate onset of its final asymptotic hierarchy.

This complex phenomenology highlights the need for a better theoretical understanding of temporal entanglement in generic Hamiltonian dynamics
at experimentally and numerically relevant timescales.

\emph{Setup. ---}
We study the real-time dynamics of a one-dimensional spin-$1/2$ quantum Ising chain in the thermodynamic limit, with transverse and longitudinal fields,
\begin{equation}
H=-J\sum_i X_iX_{i+1}-\sum_i\left(h Z_i+g X_i\right),
\label{Ising}
\end{equation}
where $X_i,$ $Z_i$ denote the corresponding Pauli matrices on site $i$. 
The model interpolates between an integrable free-fermion limit at $g=0$ and a generic ergodic regime for any finite $g$ and $h \neq 0$. Throughout this work we set $J=1$.

\begin{figure}
\centering
\includegraphics[width=1\linewidth]{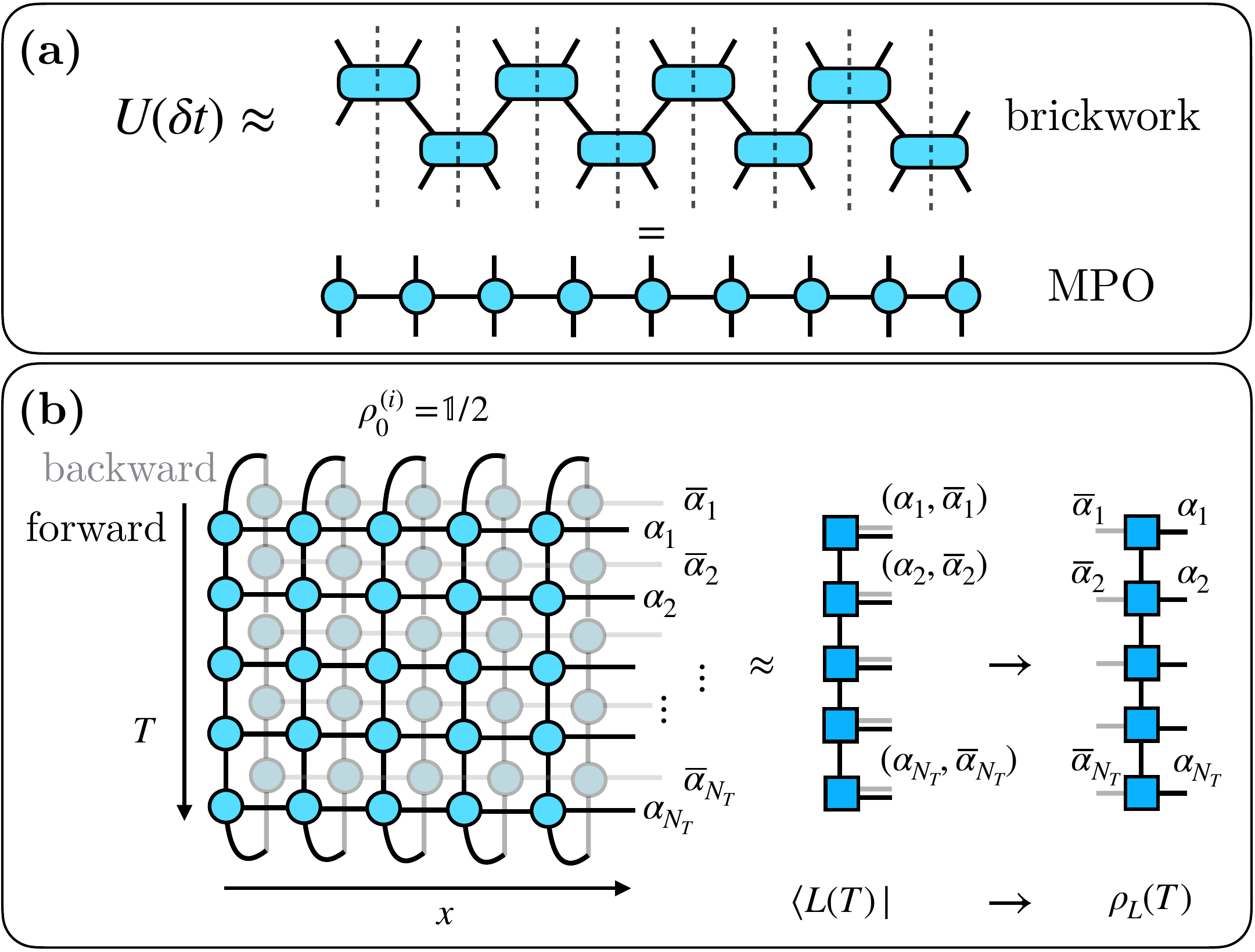}
\caption{\label{fig:sketch_IF}
\textbf{(a)} Evolution operator $U(\delta t)$, approximated by a brickwork circuit and a Matrix Product Operator. \textbf{(b)}  Influence functional for infinite temperature, represented as a density matrix, $\rho_L(T)$, together with its vectorized version, $\bra{L(T)}$.
}
\end{figure}

Time evolution can be represented as a spatio-temporal tensor network (TN)~\cite{cerezo2025}, e.g., via Trotter  and SVD decompositions of the evolution operator [Fig.~\ref{fig:sketch_IF}(a)]. For an initial state $\rho_0$, the time-evolved $\rho(T)=U(T)\rho_0 U^\dagger(T)$ can then be seen as a folded TN, pairing forward and backward evolution contours. Taking a cut along the time direction and tracing out the left (right) half-chain defines the left (right) influence functional [see Fig.~\ref{fig:sketch_IF}(b)]~\cite{feynman1963,sonner2021aop,lerose2021,ye2021}. 
The structure of the TN guarantees that, when reshaped as matrices with ket (bra) indices on the forward (backward) contour, the influence functionals are positive operators that define natural (unnormalized) density matrices for the temporal degrees of freedom and encode the temporal correlations of the system up to time $T$~\cite{cotler2018,dowling2024,odonovan2025}.
Due to the reflection symmetry of the Ising chain, both density matrices coincide, and we will focus only on the left one $\rho_L(T)$.

Our goal is to characterize $\rho_L(T)$ using Rényi entropies and temporal mutual information for different partitions of the temporal degrees of freedom. These quantities probe both the overall growth of temporal entanglement and the structure of correlations across different times, providing a detailed characterization of dynamical regimes.
Throughout the work, we focus on an infinite-temperature initial state $\rho_0=\frac{1}{\cal D} \mathbb{1}$, with ${\cal D}$ the dimension of the Hilbert space and $\mathbb{1}$ the identity operator. In that case, the full state is stationary, eliminating conventional thermalization effects and isolating the growth of temporal correlations as the sole nontrivial dynamical process.

To construct the TN, we use a second-order Trotter decomposition of the evolution operator with time step $\delta t=0.1$, and split the two-site gates using a symmetric SVD~\cite{mcculloch2007,pirvu2010matrix}. 
This yields the translationally invariant TN depicted in \cref{fig:sketch_IF}. Upon vectorization, $\rho_L(T)$ can equivalently be seen as a pure state $\bra{L(T)}$ [see \cref{fig:sketch_IF}(b)].
We use transverse contraction algorithms~\cite{banuls2009,frias2022,lerose2023overcoming,carignano2024a,carignano2026} to efficiently approximate $\bra{L(T)}$ as a Matrix Product State (MPS) with maximum bond dimension $D_{\max}=800$, enough to ensure convergence of the reported results up to $T=30$.

\emph{Temporal Rényi entropies. ---}
\begin{figure}
\centering
\includegraphics[width=0.97\linewidth]{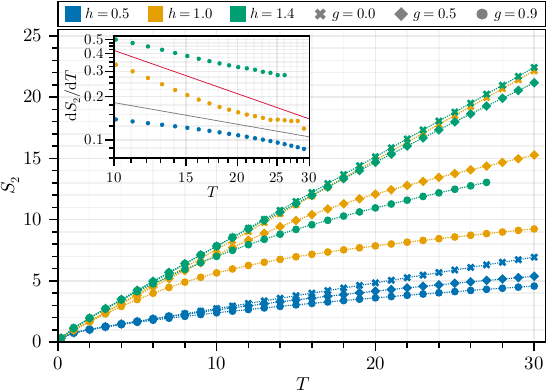}
\caption{\label{fig:S2}
Rényi-2 entropy as a function of time for representative integrable and non-integrable parameters.
Inset:
long-time behavior of $\dv*{S_2}{T}$ in a log-log plot. The solid gray and red lines displaying, respectively, the derivatives of $\sqrt{t}$ and $\log{t}$ have been included for comparison.
}
\end{figure}
We characterize temporal correlations through the Rényi entropies
\begin{equation}
    S_n(T)=\frac{1}{1-n}\log\frac{\mathrm{Tr} \rho_L^n}{(\mathrm{Tr} \rho_L)^n} \,.
    \label{renyi}
\end{equation}
We will start considering $n=2$. Using the equivalence between the density matrix $\rho_L$ and its vectorized form $\bra{L}$, we have that
\begin{equation}
S_2(T)=-\log{ \frac{\langle L|L\rangle }{ \langle L|\mathbb{1}\rangle \langle \mathbb{1}|L \rangle}} \ ,
\label{LL/LI}
\end{equation}
where the factor $\langle L|\mathbb{1}\rangle={\rm Tr} \rho_L$ ensures correct normalization. 
With our definition, $\langle L|\mathbb{1}\rangle$ is exactly exponential in $T$, a property we use to control the precision of our numerical results (see Sec.~\smref{SM:TN_details} of the Supplementary Material (SM)~\cite{SM}).
Notice that standard notions of temporal and generalized temporal entanglement are based on this vectorized form~\cite{banuls2009,hastings2015,foligno2023,carignano2024a}.

In the integrable case, the behavior is expected to agree with the quasiparticle picture~\cite{calabrese_2005}, which implies a linear growth of $S_2$ after a brief transient period. On the other hand, generic non-integrable dynamics is often modeled as a random quantum circuit, for which the membrane picture predicts a linear growth for $S_2$~\cite{foligno2023}.
However, when random circuits are supplemented with a conservation law, which makes them closer to Hamiltonian dynamics, the spatial Rényi entropies with $n>1$  are predicted to exhibit a diffusive behavior~\cite{rakovszky2018,rakovszky2019,znidaric2020,rakovszky2021,huang2020a,huang2022}, increasing as $\sqrt{t}$. The behavior of temporal entropies has been argued to mirror that of spatial entropies with additional logarithmic corrections~\cite{ippoliti2022}, so that we could expect these terms to appear in the non-integrable regime.

Figure~\ref{fig:S2} displays our results for the evolution of $S_2$ in 
the Ising chain~\eqref{Ising}. 
In the integrable regime ($g=0$), $S_2$ displays an almost immediate linear growth, in quantitative agreement with the quasi-particle picture.
In contrast, breaking integrability induces a persistent sublinear growth.
Even though we observe this behavior in all the non-integrable cases, how fast the curves evolve and deviate from the integrable ones depends strongly on the Hamiltonian parameters, with larger longitudinal field amplifying the deviation.

From the arguments above, we may expect both logarithmic and $\sqrt{t}$ terms in the long-time regime. 
A systematic analysis of our data indicates that in the time window explored, none of these forms is the dominant behavior (see Sec.~\smref{SM:sublinear} of the SM~\cite{SM}).
This is illustrated in the inset of \cref{fig:S2}, showing the finite-difference derivative of $S_2$ as a function of time for the non-integrable cases with the fastest deviation ($g=0.9$).
We conclude that there is a long transient where other terms can be contributing significantly. 
The asymptotic behavior has not 
fully emerged within the time scales accessible to our simulations, even in the examples with a faster evolution. The evaluation of higher Rényi entropies reveals the same pattern  (see Sec.~\smref{SM:higher} of the SM~\cite{SM}). 
We thus observe a deviation from generic models of ergodic dynamics along remarkably long time scales.

\begin{figure}
    \centering
    \includegraphics[width=0.98\linewidth]{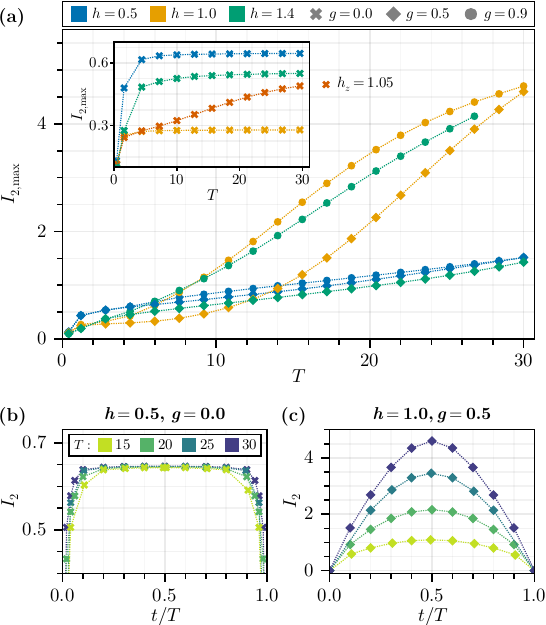}
    \caption{\label{fig:MI}
    \textbf{(a)} Maximum of the temporal mutual information $I_2$ for the bipartition
    $A=[0,t]$ and $B=[t+\delta t,T]$ for 
     non-integrable and integrable dynamics (inset).
$I_2$ as function of $t/T$ for integrable \textbf{(b)} and non-integrable  \textbf{(c)} dynamics at different times $T$. 
    }
\end{figure}

\emph{Mutual information. ---}
The correlations between different temporal intervals, which we can probe using the Rényi-2 mutual information, provide a more detailed insight into the structure of the influence matrix: whereas the entropies discussed above are related to its degree of mixedness, temporal correlations relate to memory effects in the effective bath and directly impact its approximability by Matrix Product Operators (MPOs) \cite{wolf2008}.

We consider first the bipartition $A=[0,t]$ and $B=[t+\delta t,T]$.
Tracing out the temporal degrees of freedom in $B$ returns, up to normalization, the influence matrix for the time span covered by $A$ (see Sec.~\smref{SM:TN_details} of the SM~\cite{SM}), so that
\begin{equation}
S_2(A;T)=S_2(t). \ 
\label{s2A}
\end{equation} 
Similarly, the $t \to T-t$ symmetry of our setup implies $S_2(B;T)=S_2(T-t)$.
The mutual information of the bipartition $A:B$ can be thus written as 
\begin{equation}
I_2(A\!:\!B\,;T)=S_2(t)+S_2(T-t)-S_2(T).
\label{mutual}
\end{equation}

In the integrable regime, the linear behavior of $S_2$ shown in Figure~\ref{fig:S2} implies a strict area law for the temporal mutual information.
This is confirmed by the inset of Fig.~\ref{fig:MI}(a), where we plot the maximum of the mutual information, for the symmetric bipartition, $t=T/2$, with the infinite-temperature initial state.
The saturation of $I_2$ with block size is in general rapidly achieved, experiencing a slowdown close to the critical value of the 
magnetic field at equilibrium, $h=1$, as illustrated by the $h=1.05$ line~\cite{hohenberg1977}. 

The strong difference between integrable and non-integrable regimes is clearly appreciated by comparing Figure~\ref{fig:MI}(b) and \ref{fig:MI}(c). Indeed, 
once integrability is broken, the mutual information is no longer seen to saturate for any value of the transverse or longitudinal magnetic fields within the accessible times. Figure~\ref{fig:MI}(a) shows that its maximum undergoes a regime of apparent linear growth at mesoscopic time scales.
Visibly, $(h=1,g=0.9)$  has exited this phase at $T \sim 20$, while the slowest curves do not show signs of bending down in the accessed window.
A linear growth, however, cannot be sustained over time, since it would result in unphysical negative values of the entropy, as implied by \eqref{mutual}. 
This further demonstrates the presence of a transient regime at mesoscopic time scales, in which correlations between different times rapidly build up.

\emph{Correlations at finite temporal distance. ---} 
To further explore the structure of temporal correlations, we compute the Rényi-2 
mutual information between two disjoint intervals of temporal extension $t_{l}$ separated by a distance $\Delta t$, $I_2(t_{l},\Delta t)$. Given the symmetry of our setup, it 
does not depend on the absolute location of the intervals or the total time $T$. 
Notice that 
$I_{2,\max}$ plotted in Fig.~\ref{fig:MI}(a) corresponds to the limiting case: $\Delta t=0$, $t_{l}=T/2$.

\begin{figure}[t]
    \centering
    \includegraphics[width=1.0\linewidth]{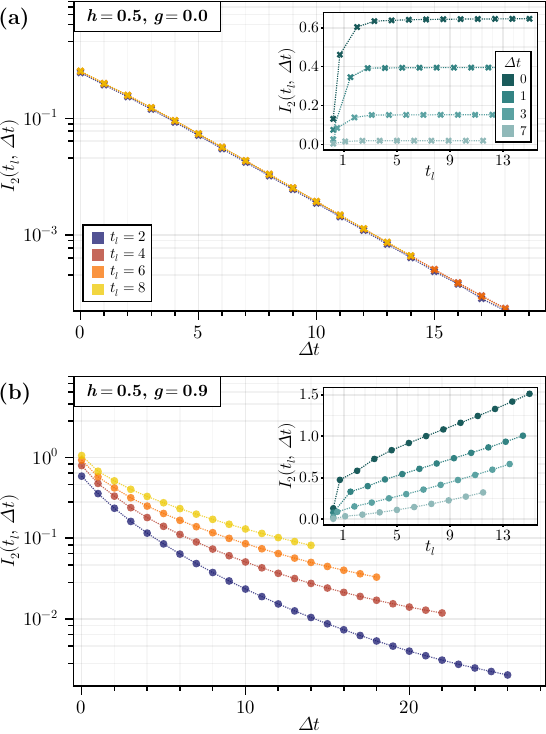}
    \caption{
    Mutual information $I_2$ between two temporal blocks of length $t_l$ as a function of their separation $\Delta t$, for
    \textbf{(a)} integrable ($h=0.5$)
    and \textbf{(b)} non-integrable ($h=0.5$, $g=0.9$) dynamics.
  }
    \label{fig:MI_2blocks}
\end{figure}

Figure~\ref{fig:MI_2blocks} highlights the distinct features of $I_2(t_{l},\Delta t)$ in different dynamical regimes for representative parameters. The integrable case, illustrated in panel (a), behaves as expected from a quasiparticle picture. For fixed interval length $t_{l}$, correlations decay exponentially with the separation $\Delta t$. Fixing the interval $\Delta t$ (see inset), the mutual information becomes independent of the block size, evidencing an area law.
In the non-integrable case, correlations between blocks of fixed size instead decay with distance at a markedly slower rate, now exhibiting a strong dependence on the block size, as shown in panel (b). At fixed separation (see inset),   $I_2(t_{l},\Delta t)$ 
grows without signs of saturation over a wide interval range, generalizing the $\Delta t=0$ results in Fig.~\ref{fig:MI}. 

The slower decrease with $\Delta t$ in the non--integrable case seems to be better described by an exponential decay. However, a systematic analysis shows that it is not possible to unambiguously discriminate the functional form within the simulated time window (see Sec.~\smref{app:fits-mutual-info-decay} of the SM~\cite{SM}), suggesting again that the asymptotic behavior has not yet been fully reached.

\begin{figure}
    \centering
    \includegraphics[width=0.97\linewidth]{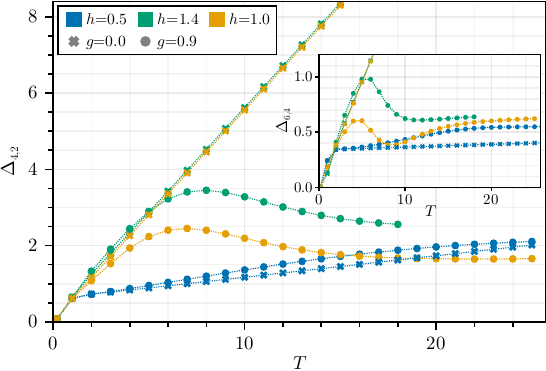}
    \caption{\label{fig:S2vec}
Evolution of the relative Rényi scalings $\Delta_{4,2}$ and $\Delta_{6,4}$ (inset) for integrable and non-integrable dynamics.
    }
\end{figure}

\emph{Relative Rényi scalings. ---}
\label{sec:higherSn}
It is interesting to compare the scaling of different Rényi entropies, for which we introduce, for integer $n>m$, the quantities
\begin{equation}
\Delta_{n,m}  =-\log\frac{\mathrm{Tr} \rho_L^n}{{(\mathrm{Tr} \rho_L^m)}^{\frac{n}{m}}}
\ ,
    \label{eq:d34}
\end{equation}
We consider $m \geq 2$, since $\Delta_{n,1}=(n-1)S_n$. 

While in the integrable case these quantities grow linearly with time, in the non-integrable regime they display a pronounced non-monotonic behavior. The evolution of $\Delta_{4,2}$ is shown in Fig.~\ref{fig:S2vec}, where we observe that after an initial growth it reaches a maximum and subsequently decreases. We have analyzed all quotients \eqref{eq:d34} with $n$ up to 6, finding analogous results. As $m$ increases, the maximum is attained at slighly earlier times 
(see inset; also Sec.~\smref{SM:Deltas} of the SM~\cite{SM} and Ref.~\cite{hurley2009comparing} therein).

This seemingly surprising behavior is qualitatively captured by a simple toy model where the (normalized) spectrum of $\rho_L$ features a single dominant eigenvalue $\lambda$ alongside $N$  degenerate ones, with $N$ growing exponentially in time. For example, $\lambda=e^{-\sqrt{D t}}$ would be associated to a diffusive behavior, but any sublinear dependence of the exponent with time is equally valid for our arguments.  When $\lambda$ starts dominating over the collective contribution of the degenerate eigenvalues, the corresponding Renyi scaling attains a maximum, which is followed by a monotonic decay to zero.  This latter feature however disagrees with our observations: the inset of Fig.~\ref{fig:S2vec} shows that, after the maximum, $\Delta_{6,4}$ for the fastest evolving example $(h=1,g=0.9)$ reaches a minimum and starts growing again. Including two slowly decaying eigenvalues which at some point exchange the dominant role, also that behavior can be reproduced (see Sec.~\smref{SM:toy} of the SM~\cite{SM}). 
This provides a valuable insight into the transient regime, linking it to significant changes in  the eigenvalue structure of $\rho_L$, well into mesoscopic timescales.

Notice that $\Delta_{2n,2}$ equals $(n-1)$ times the Rényi$-n$ entropy 
of  the pure state $\bra{L}/\sqrt{\langle{L}\ket{L}}$ with respect to the forward--backward bipartition, measuring thus entanglement between contours. Contrary to the scalings \eqref{eq:d34} with $m\!>\!2$, these entropies exhibit a monotonic decay after the maximum  in the time window numerically accessible. Hence, the rapid  increase of temporal correlations detected by the mutual information in the transient regime, appears to be accompanied by a decrease of the entanglement between forward and backward contours. Its potential implications for the efficiency of  folding algorithms in non-integrable systems deserves further study~\cite{Schuch_2008}.

\emph{Summary and Discussion. ---}
In this work,
we have characterized the structure of temporal correlations generated by interacting quantum dynamics through the influence functional of a half-infinite system at infinite temperature. In integrable regimes, the temporal entropy growth follows the quasiparticle picture as expected.

In striking contrast, generic non-integrable Hamiltonian dynamics exhibits persistent deviations from the random-circuit paradigm even when enriched with conserved quantities. At the time scales we are able to simulate, we do not see neither a behavior fully consistent with diffusion nor a well defined asymptotic regime. A toy model with a few leading eigenvalues slowly evolving over the exponentially large remaining part of the spectrum, is able to qualitatively explain the main features we observe. 

Concluding, these results demonstrate that ergodic Hamiltonian dynamics develops a rich intermediate structure of temporal entanglement. 
Deviations between random circuits and Hamiltonian dynamics at the level of generalized temporal entropies in Loschmidt-echo scenarios we have been also observed in~\cite{carignano2025}.
Our findings suggest limitations of the current universality paradigms for non-equilibrium quantum dynamics, and 
have direct implications for the complexity and simulability of influence functionals in interacting systems. In future work, we aim to extend our analysis to initial pure states, and to deepen the connection between the structure of temporal correlations and the quasiparticle picture within the integrable regime \cite{cerezo2026}.

\emph{Aknowledgments. ---} We thank Jacopo De Nardis and Mar\'{\i}a Cea for the collaboration and discussion on related subjects.
We acknowledge the support from the Proyecto Sinérgico CAM Programa TEC-2024/COM-84 QUITEMAD-CM, the CSIC Research Platform on Quantum Technologies PTI-001, from the Grant TED2021-130552B-C22 funded by MCIN/AEI/10.13039/501100011033 and by the ``European Union NextGenerationEU/PRTR'', from the grant PID2024-160172NB-I00 funded by MICIU/AEI/10.13039/501100011033 and by FEDER,UE,  from the Spanish Agencia Estatal de Investigaci\'on through “Instituto de F\'{\i}sica
Te\'orica Centro de Excelencia Severo Ochoa CEX2020-001007-S” and from Grant PID2021-127968NB-I00 funded by MCIN/AEI/10.13039/501100011033 

SCR thanks support from FPU23/02915 scholarship from the MCIU.
JTS acknowledges support from the Programa Fundamentos FBBVA through the grant EIC24-1-17304, and the support by the Ministry for Digital Transformation and of Civil Service of the Spanish Government through the QUANTUM ENIA project call - Quantum Spain project and by the European Union through the Recovery, Transformation and Resilience Plan - NextGenerationEU within the framework of the Digital Spain 2026 Agenda. The funding for these actions/grants and contracts comes from the European Union’s Recovery and Resilience Facility-Next Generation, in the framework of the General Invitation of the Spanish Government’s public business entity Red.es to participate in talent attraction and retention programs within Investment 4 of Component 19 of the Recovery, Transformation, and Resilience Plan.
ABC is supported by Grant CQS2301001 from the Project Quantum ENIA. The funding for this grant comes from the Plan de Recuperación, Transformación y Resiliencia en el marco de la Agenda España Digital 2025, from the Ministerio de Asuntos Económicos y Transformación Digital. ABC is also supported by Grant MMT24IFF-01.
SC acknowledges his AI4S fellowship within the “Generación D” initiative by Red.es, Ministerio para la Transformación Digital y de la Función Pública, for talent
attraction (C005/24-ED CV1), funded by NextGenerationEU through PRTR. We also acknowledge PID2023-151384NB-I00 financed by MCIU/AEI/10.13039/501100011033 and FSE+.

MCB acknowledges funding from the Deutsche Forschungsgemeinschaft (DFG, German Research Foundation) under Germany's Excellence Strategy -- EXC-2111 -- 390814868, and through the Research Unit FOR 5522/1, Grant No. 499180199.

\bibliography{References.bib}

\ifdefined\IsMainDocument
\ifdefined\IsMainDocument
    \clearpage
    \onecolumngrid
    \begin{center}
    \textbf{\large Supplemental Material -- Mesoscopic Regimes of Temporal Entanglement in Ergodic Quantum Systems}
    \end{center}
    
    \setcounter{section}{0}
    \setcounter{equation}{0}
    \setcounter{figure}{0}
    \setcounter{table}{0}
    \setcounter{page}{1}

    \newcommand{\mainref}[1]{\cref{#1}}
\else

    \documentclass[%
    a4paper,
    aps,
    prl,
    onecolumn,
    nofootinbib,
    amsmath,amssymb,
    superscriptaddress,
    10pt,
    floatfix,
    ]{revtex4-2}
    
    \usepackage[a4paper,centering,hmargin=2.25cm,vmargin=1.75cm]{geometry}
    
    \usepackage{graphicx}
    \usepackage{amsfonts,amsmath}
    \usepackage{mathtools}
    \usepackage{amsmath}
    \usepackage{xcolor}
    \usepackage{bbold}
    \usepackage{physics}
    \usepackage{braket}
    \usepackage{ulem}
    \usepackage{microtype}
    \usepackage{xr} 
    \usepackage{afterpage}
    \usepackage[colorlinks=true,citecolor=blue,linkcolor=blue,urlcolor=blue,filecolor=black]{hyperref}
    \usepackage[capitalize]{cleveref}

    \newcommand{\mainref}[1]{\textcolor{black}{\cref*{#1}}}
    \externaldocument{letter_mesoscopic}
    
    \let\emph\relax
    \DeclareTextFontCommand{\emph}{\itshape}
    
    \def\bbI{\mathbb{I}}
    \def\bbO{\mathcal{O}}
    \def\bbQ{\mathcal{Q}}
    \def\TM{\mathcal{T}}
    \def\tmE{\mathbb{E}^T}
    \def\tmEp{\mathbb{E}^{T+1}}
    
    \def\rank{\mathcal{R}}
    \def\Lt{\Lambda^{s}}  
    \def\Lb{\Lambda^{o}}
    \def\Lbt{{\bar\Lambda}^{s}}  
    \def\Lbb{{\bar\Lambda}^{o}}
    \def\Ts{{\mathcal T}}

    \newcommand{\Eq}[1]{{Eq.~({\ref{#1}})}}
    \newcommand{\Fig}[1]{{Fig.~{\ref{#1}}}}
    
    \newcommand{\beq}{\begin{equation}}
    \newcommand{\eeq}{\end{equation}}
    
    \begin{document}

    \title{Supplemental Material -- Mesoscopic Regimes of Temporal Entanglement in Ergodic Quantum Systems}
    
    \author{Sergio Cerezo-Roquebrún}
    \affiliation{Instituto de Física Teórica UAM/CSIC, C/ Nicolás Cabrera 13--15, Cantoblanco, 28049 Madrid, Spain}
    \author{Jan Thorben Schneider}
    \affiliation{Institute of Fundamental Physics IFF-CSIC, C/ Serrano 113b, Madrid 28006, Spain}%
    \author{Stefano Carignano}
    \affiliation{Barcelona Supercomputing Center, 08034 Barcelona, Spain}
    \author{Aleix Bou-Comas}
    \affiliation{Institute of Fundamental Physics IFF-CSIC, C/ Serrano 113b, Madrid 28006, Spain}%
    \author{Mari Carmen Bañuls}
    \affiliation{Max-Planck-Institut fur Quantenoptik, Hans-Kopfermann-Str.~1, D-85748 Garching, Germany}
    \affiliation{Munich Center for Quantum Science and Technology (MCQST), Schellingstr.~4, D-80799 München, Germany}
    \author{Esperanza López}
    \affiliation{Instituto de Física Teórica UAM/CSIC, C/ Nicolás Cabrera 13--15, Cantoblanco, 28049 Madrid, Spain}%
    
    \author{Luca Tagliacozzo}
    \affiliation{Institute of Fundamental Physics IFF-CSIC, C/ Serrano 113b, Madrid 28006, Spain}
    \email{luca.tagliacozzo@iff.csic.es}

    \maketitle

\fi

\setcounter{secnumdepth}{2}
\renewcommand{\thesection}{S\arabic{section}}
\renewcommand{\theequation}{S\arabic{equation}}
\renewcommand{\thefigure}{S\arabic{figure}}
\renewcommand{\thetable}{S\arabic{table}}

\ifdefined\IsMainDocument
\else
    {
      \hypersetup{linkcolor=black}
      \tableofcontents
    }
    \vspace{1cm}
    
\fi

\section{TN details}
\label{SM:TN_details}

\begin{figure}[h]
\centering
\includegraphics[width=14cm]{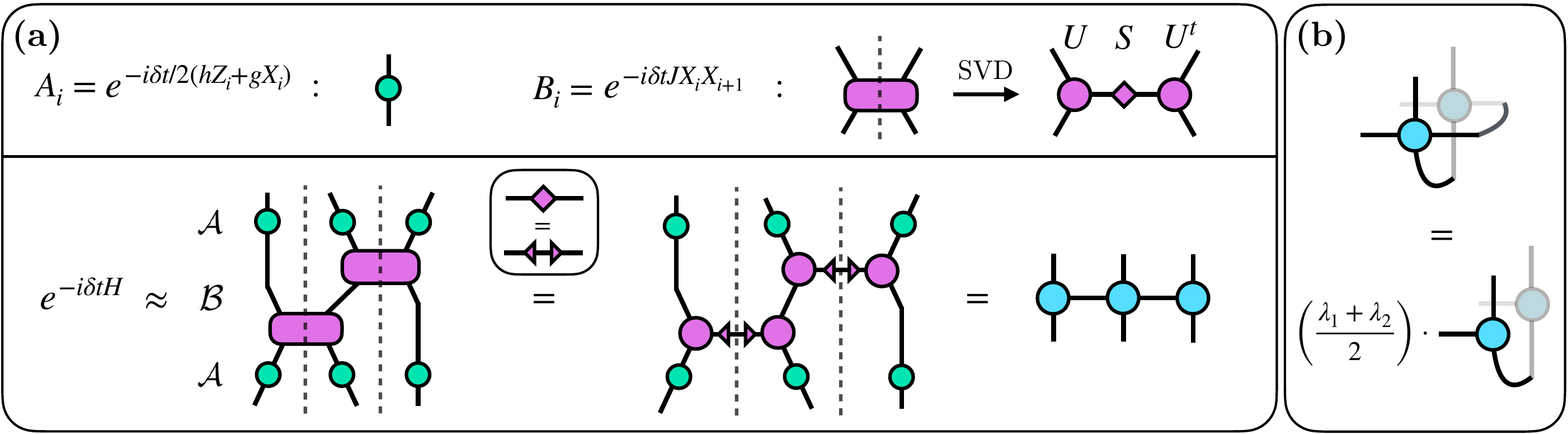}
\caption{Building-blocks for the time evolution and recursion relations of the influence functional. \textbf{(a)} Top: elementary 1-qubit $A_i$ and 2-qubit $B_i$ gates. The latter is decomposed through a symmetric SVD splitting the degrees of freedom of the qubit $i$ and $i+1$. Bottom: construction of the MPO representation of $e^{-i\delta t H}$. \textbf{(b)} Relation satisfied by the 4--leg and 3--leg tensors of the MPO used for the evolution.}
\label{fig:TN_details}
\end{figure}

Our numerical results are based on the TN description of the influence functional shown in \mainref{fig:sketch_IF}.
The Hamiltonian evolution of the Ising model is approximated by a second--order Trotter--Suzuki expansion, $e^{-i \delta t H}\approx {\cal A} {\cal B} {\cal A} $, where ${\cal A}=\prod_i A_i$ is the product of single-qubit rotations induced by the transverse and longitudinal magnetic fields, $e^{-i\delta t/2(hZ_i + gX_i)}$, and ${\cal B}=\prod_i B_i$ that of the entangling gates $e^{-i \delta t  J X_i X_{i+1}}$. By grouping the indices associated with each qubit the $B_i$ gates are further split via an SVD, 
with singular values $\lambda_1=2\cos J \delta t$ and $\lambda_2=2\sin J \delta t$. The 3-- and 4--leg tensors building the spatio-temporal TN are pictorially defined in \cref{fig:TN_details}(a). Additionally, for this model, it can be seen that the partial contraction of the 4--leg tensor with its adjoint reduces to that of the 3--leg tensor, up to a factor of $(\lambda_1+\lambda_2)/2$ (see \cref{fig:TN_details}(b)). 

With the above conventions, the trace of the influence functional, represented as the density matrix $\rho_L(T)$, 
grows exponentially with time as
\begin{equation}
     {\rm Tr} \, \rho_L(T)=\left( \frac{\lambda_1+\lambda_2}{2}\right)^{\frac{T}{\delta t}} \ .
     \label{tr}
 \end{equation}

 We can apply the relation in Fig.~\ref{fig:TN_details}(b) also to only a partial trace. In particular, if we 
 consider the reduced density matrix obtained by tracing $\rho_L(T)$ over a subsystem $B=[t+\delta t,T]$, we have
\begin{equation}
     {\rm Tr}_B \, \rho_L(T)= \left( \frac{\lambda_1+\lambda_2}{2} \right)^{\frac{T-t}{\delta t}} \, \rho_L(t) \ ,
 \end{equation}
where $\rho_L(t)$ is the influence functional at time $t$. This relation implies that the Rényi entropies of the reduced density matrix coincide with those of the influence functional at the earlier time $t$. Due to the 
the symmetry of the infinite temperature
TN under $t \to T-t$, an analogous relations holds when tracing over $A = [0, t ]$.

Because \cref{tr} is exact, we can use it to monitor the truncation error in our numerical results. If the truncation error is sufficiently small, the trace of the computed influence functional will grow exponentially in time with the exact rate, namely $\frac{1}{T}\log( {\rm Tr} \rho_L)=\frac{1}{T}\log \bra{L}\mathbb{1}\rangle=\frac{1}{\delta t} \log \left( \frac{\lambda_1+\lambda_2}{2} \right)$. Thus, to ensure the quality of the data used for the fits, we require that deviations with respect to the predicted value stay within $0.1\%$. 

 Another potential source of numerical errors is the finite Trotter step used in our simulations.
For most of the reported results, we have used a fixed time step $\delta t=0.1$. To ensure that the observed phenomenology is not an effect of this choice, and are in fact characteristic of the Hamiltonian evolution, we have compared the results with a smaller step $\delta t=0.05$. We found that for all non-integrable cases reported in the main text, the deviation in $S_2$ between both time steps is below $5\cdot 10^{-3}$ for all times $T\leq 30$ for bond dimensions $D=500,\, 800.$
 
\section{Fits of \texorpdfstring{$S_2$}{S2} in the non-integrable regime}
\label{SM:sublinear}

\begin{figure}[h]
    \centering
    \includegraphics[width=\linewidth]{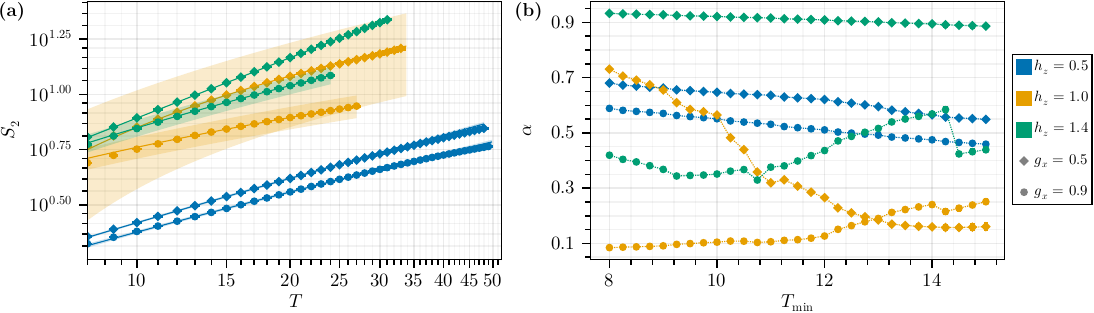}
    \caption{\label{fig:sublinear}
    $S_2(T)$ in the non-integrable regime.
    \textbf{(a)} Log-log plot of $S_2(T)$ (markers) with an error estimate inferred from the difference between the simulation results for $\chi_m = 800$ and $\chi_m = 500$; consistently smaller than the marker size.
    The line marks the average fit of a power-law model while varying the lower threshold of fitted data. The error band reflects the propagated error estimate over the ensemble of fits and the fit errors of the parameters.
    \textbf{(b)} The exponent of the power-law model $\alpha$ inferred from fits to $S_2$ plotted against $T_\mathrm{min}$ marking the lower threshold of the fitted window.
    }
    
\end{figure}

In this section, we present the analysis of our late-time data for $S_2$ and its behavior 
in the non-integrable regime. We plot the data for $S_2$ once more as markers in \cref{fig:sublinear}(a).
The slow growth of $S_2$ over the observed window implies that $\sqrt{T}$ and $\log T$ contributions cannot be unambiguously separated by a fit over the available times.
Fits of the form $S_2 = A \sqrt{T} + C$ and $S_2 = A' \log{T} + C'$ can approximate the observed behavior, but both yield highly structured residuals, which points towards an incomplete description of the functional form.
To investigate the asymptotic behavior of the entropy, we thus fit a free power law model $S_2(T) = A \, T^{\alpha} + C$, to different windows [$T_{\mathrm{min}}$, $T_{\max}$].
We fix $T_{\max}$ to the latest simulated time for which the computed value of $\log(\bra{L}\mathbb{1}\rangle)/T$ is within $10^{-3}$ of the exact one, and vary the lower limit $T_\mathrm{min} \in \left\{8, 8.25, \ldots, 15\right\}$. This gives us an ensemble of power-law fits.
The ensemble average power-law model is plotted in \cref{fig:sublinear}(a) as the solid line.
The ensemble error estimate, which is inferred from ensemble standard deviation, and the fit uncertainties, is plotted as the band.
The fitted exponent $\alpha$ of each ensemble member is shown in \cref{fig:sublinear}(b).
If the accessible long-time behavior were described by a single power-law model, the exponents would not depend strongly on the fitted window. As one can observe in \cref{fig:sublinear}(b), however, some of the exponents drift with changing the lower boundary of the fitted windows, $T_\mathrm{min}$.
The clear drift for most parameters is consistent with a transient regime where the asymptotic behavior has not settled in yet.
In \cref{tab:exponents}, we print the arithmetic mean of the parameters and their propagated error estimate.
\begin{table}[ht]
\centering
\setlength{\tabcolsep}{9pt}
\renewcommand{\arraystretch}{1.5}
\begin{tabular}{c|c |c}
 & $g = 0.5$ & $g = 0.9$ \\
\hline
$h = 0.5$ & $0.616 \pm 0.0396$ & $0.521 \pm 0.0404$ \\\hline
$h = 1.0$ & $0.245 \pm 0.100$  & $0.174 \pm 0.0551$ \\\hline
$h = 1.4$ & $0.910 \pm 0.0142$ & $0.431 \pm 0.0783$ \\
\end{tabular}
\caption{Power-law exponents $\alpha$ from fits of $S_2 = T^\alpha +C$ with $T_\mathrm{low} \in \left\{8, 8.25, \ldots, 15\right\}$. The value is given by the arithmetic mean of the fit values, while the error is propagated of the fit error and the standard deviation of the fit values.}
\label{tab:exponents}
\end{table}
We observe that diffusive behavior, which would be marked by an exponent of $1/2$, cannot fully capture the data, and only one data set is consistent with this behavior ($h=0.5$, $g = 0.9$).

\section{Higher Rényi entropies}
\label{SM:higher}

To compute the higher Rényi entropies, we leveraged the MPO structure of $\rho_L$ to iteratively multiply and compress its powers, limiting the maximum bond dimension of the resulting MPOs to $2D$, where $D$ is the bond dimension used for $\rho_L$. Due to memory requirements, we have restricted ourselves to $D=200$. As a result, the norm-preserving criterion described in \cref{SM:TN_details} halts the evolution at earlier times than those reached for $S_2$.

\begin{figure}
\centering
\includegraphics[width=0.9\linewidth]{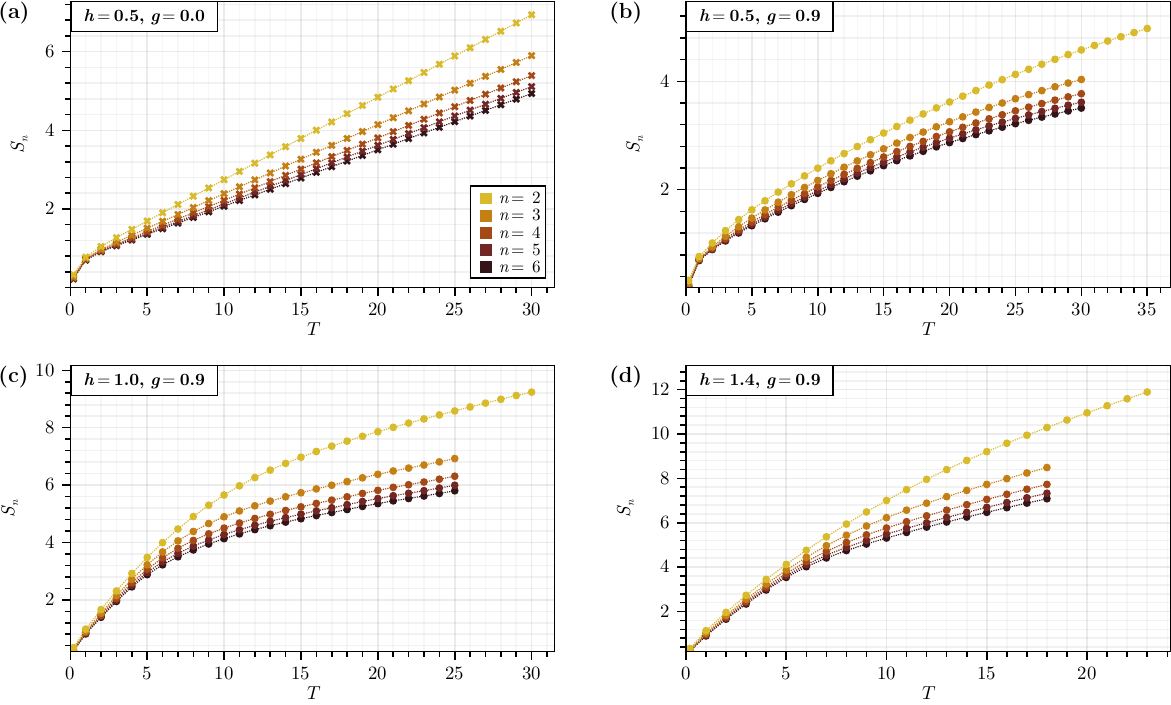}
\caption{Rényi entropies $S_n$ of the influence functional $\rho_L$ as a function of time, for $2\leq n \leq 6$, for some representative \textbf{(a)} integrable, and \textbf{(b)}--\textbf{(d)} chaotic  evolutions.}
\label{fig:higher-renyis}
\end{figure}

For completeness, we report our results for $S_n$, with $2\leq n \leq 6$, in \cref{fig:higher-renyis}. In the integrable regime, all R\'enyi entropies grow linearly with time in consonance with the quasiparticle picture, see panel \textbf{(a)}. On the other hand, once integrability is broken, the growth pattern becomes sublinear, as shown in panels \textbf{(b)}-\textbf{(d)}. 
Notice that \cref{fig:higher-renyis} illustrates the monotonicity of the R\'enyi entropies with the index $n$. 
 
\section{Mutual information of separated intervals in the non-integrable regime}\label{app:fits-mutual-info-decay}

\begin{figure}
    \centering
    \includegraphics[width=0.9\linewidth]{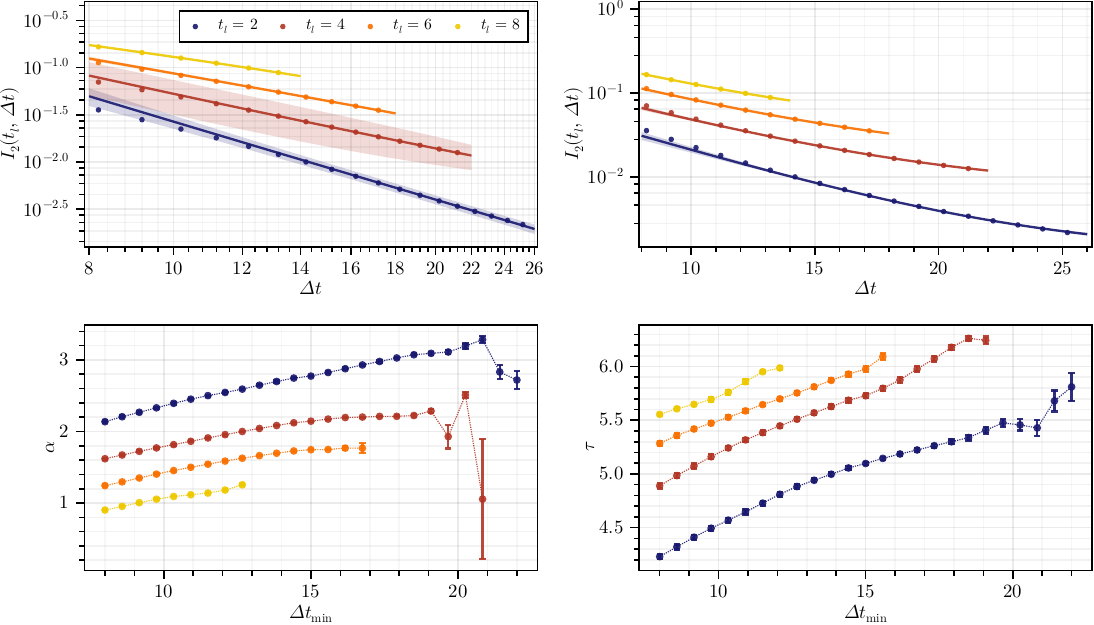}
    \caption{\label{fig:fits-mutual-info}
    Fits modelling the long-time decay of $I_2(\Delta t, t_l)$ as a function of $\Delta t$. Left column: model is $I_2 = A_\alpha \Delta t^{-\alpha} + C_\alpha $, in the right column model is $I_2 = A_\tau \exp(\Delta t/\tau) + C_\tau$.
    First row shows the data for $I_2(\Delta t, t_l)$ in a log-log plot (left), and a log-lin plot (right), for values of $\Delta t \geq 8$.
    We fit the respective model to the data over a range of $\Delta t \in [\Delta t_\mathrm{min},...]$ up to the maximum available data. Varying the lower fit boundary gives us an ensemble of different fits.
    We plot the ensemble average fit as the solid line, and the band reflects the uncertainty propagated from fit uncertainty and ensemble standard deviation.
    Second row shows the fitted decay parameter for each ensemble member, left the power-law exponent $\alpha$, and right the exponential decay rate $\tau$.
    }
\end{figure}

In this appendix, we attempt to model the decay of mutual information of two blocks of length \(t_l\) and separation $\Delta t$ in the non-integrable regime. The raw data is shown in \mainref{fig:MI_2blocks}(b) in the main text, which shows the case for $g=0.9$ and $h=0.5$.
To better extract the long-time behavior, we fit the finite-difference derivative of the data to the analytical derivative of the model in order to get rid of any constant offset that the data presents at long times. Note that we plot only every 5th data point to better visually discern data.
As models, we employ a power-law model, $I_2 = A_\alpha \Delta t^{-\alpha} + C_\alpha $,  and an exponential model, $I_2 = A_\tau \exp(\Delta t/\tau) + C_\tau$; both decay up to a constant which we fit.
To distinguish which of the two models is preferred, we fit both models to an ensemble of data, made up of different windows of $\Delta t \in [\Delta t_\mathrm{min},...,]$.
We vary $\Delta t_\mathrm{min} \geq 8$, and fit up to the maximum available data; this gives us an ensemble of fits.
Our results are presented in \cref{fig:fits-mutual-info}, where we show the data (markers, top row) in a log-log plot (left) and in a lin-log plot (right).
The left column is testing for the power-law model, while the right column is testing for the exponential decay model.
In the lower row, we plot the fitted decay parameter for each ensemble member, left the power-law exponent $\alpha$, and right the exponential decay rate $\tau$.
We observe a clear drift of the parameters with lower fit threshold $\Delta t_\mathrm{min}$,
indicating that the parameters did not yet settle to a constant regime for very late times, and corroborating the narrative that the numerically accessible regime is still transient.
To better distinguish between the two models, we compute the Akaike information criterion (AIC) for all models at the different fitted windows.
As the numerical values naturally change with the different data when fitting to different $\Delta t$ windows, we compare at each $\Delta t_\mathrm{min}$ the AIC (lower is better) over the fits being displayed in \cref{fig:fits-mutual-info}(lower row).
These win-loss counts for the exponential and power-law models are printed in \cref{tab:mutual-info-decay}.
\begin{table}[h!]
\centering
\setlength{\tabcolsep}{9pt}
\renewcommand{\arraystretch}{1.5}
\begin{tabular}{c | l | c}
$t_l$ & AIC preference for each $\Delta t_\mathrm{min}$ (power-law = 0, exponential = 1) & frequency, winner \\
\hline
2 & \texttt{1011000111111111100010011} & $16 / 25$, exponential \\
4 & \texttt{11111111110000000011000} & $12 / 23$, exponential \\
6 & \texttt{1111111111000000} & $10 / 16$, exponential\\
8 & \texttt{111100110} & $6/9$, exponential\\
\end{tabular}
\caption{\label{tab:mutual-info-decay}%
Evaluation of preferred model for decay of $I_2$ with $\Delta t$ for different $t_l$.
AIC of the fitted model (lower is better) determines which model is preferred for the given $\Delta t_\mathrm{min}$ and $t_l$. Center column shows the win-loss count for the fits over varied $\Delta t_\mathrm{min}$ shown in \cref{fig:fits-mutual-info}(lower row). Right column shows the win-loss count and the preferred model.
}
\end{table}
It is however important to note that both models include a free constant offset to be fitted which implies the mutual information may be modelled to not decay to zero.
For all models fitted, the constant offset for the exponential decay is larger $C_{\tau} > C_{\alpha}$, most pronounced for $t_l = 2$ where $C_{\alpha} = -6.7 \cdot 10^{-5} \pm  9\cdot 10^{-6}$ and $C_{\tau} = -1.25 \cdot 10^{-3} \pm 4\cdot 10^{-6}$. For these models, the late dynamics is better described by an exponential decay to a constant offset.
Nevertheless, we conclude that the long-time decay of $I_2$ with $\Delta t$ and over longer time scales of $\Delta t$ show, cf.\ the drifting fitted parameters (\cref{fig:fits-mutual-info}(lower row)), that within the simulated time window, the asymptotic behavior has not yet been reached, and an unambiguous discrimination of the two models is out of reach.

\section{Relative Rényi scalings \texorpdfstring{$\Delta_{n,m}$}{Δₙ,ₘ}}
\label{SM:Deltas}

\begin{figure}
\centering
\includegraphics[width=1\linewidth]{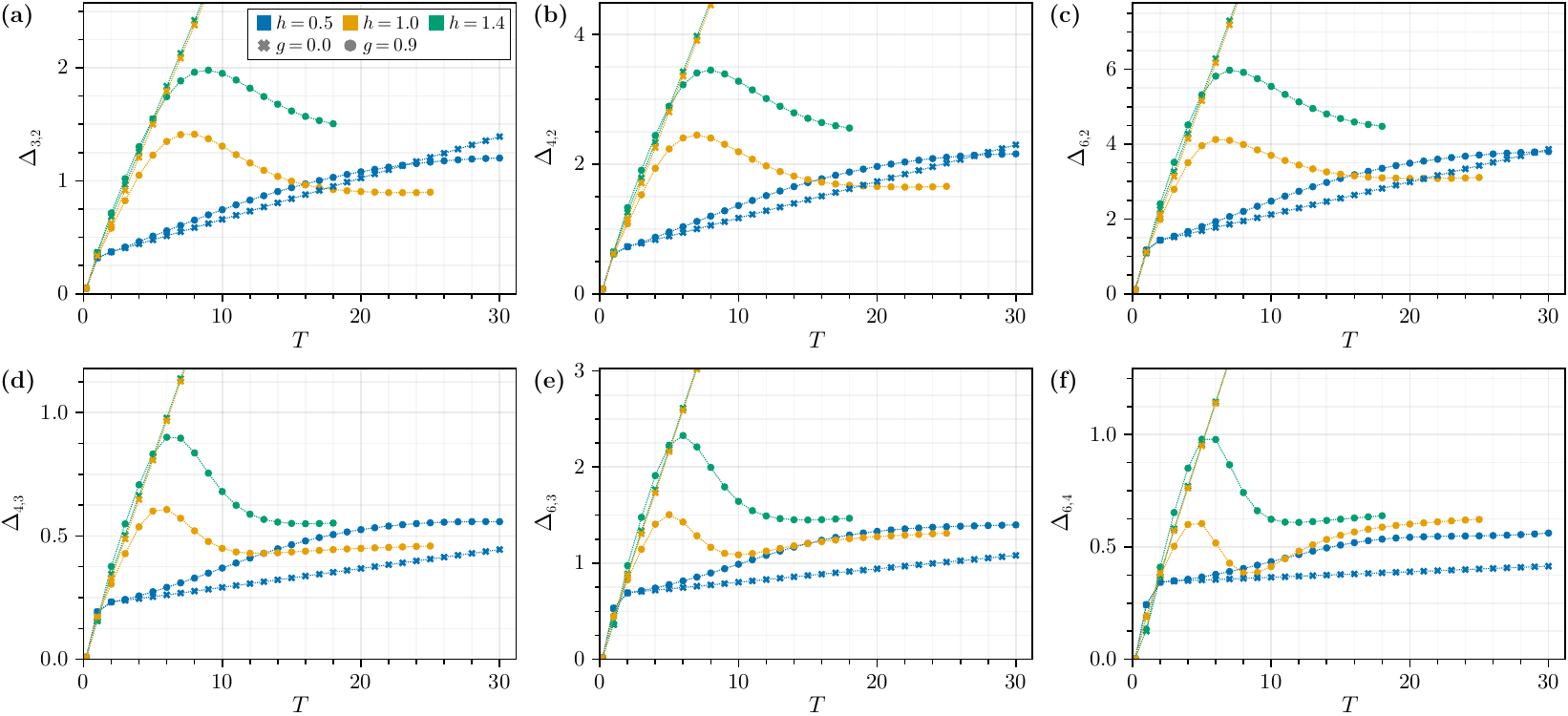}
\caption{Relative Rényi scaling $\Delta_{n,m}$ for different pairs $(n,m)$.}
\label{fig:Delta_nm}
\end{figure}

In this section, we analyze the behavior of the different R\'enyi scalings, $\Delta_{n,m}$ \mainref{eq:d34}, for $n>m\geq2$. 
These quantities can be written as a ratio between Schatten $p$--norms $\|M\|_p=[{\rm Tr} (MM^{\dagger})^{p/2}]^{1/p}$,
\begin{equation}
    \Delta_{n,m} =-\log\frac{\mathrm{Tr} \rho_L^n}{{(\mathrm{Tr} \rho_L^m)}^{\frac{n}{m}}} = - n \log \cfrac{\|\rho_L\|_n}{\|\rho_L\|_m}\,,
    \label{deltanm}
\end{equation}
generalizing the quotient $(\| \rho_L \|_4 / \| \rho_L\|_2)^4$, which has been used as a measure of sparsity \cite{hurley2009comparing} for the spectrum of a matrix. Here, sparsity means that few eigenvalues are large proportionally to the rest, i.e. the spectrum of $\rho_L$ is peaked. Thus, $\Delta_{n,m}$ is minimal and equal to zero for a rank-1 matrix, while at fixed ${\rm rank}\,\rho_L(T)=N$ it is maximal when all the eigenvalues are equal, taking the value $\Delta_{n,m}=(n/m-1)\log N$. In contrast to other sparsity measures \cite{hurley2009comparing}, the family (\ref{deltanm}) accounts only for the non-zero eigenvalues of $\rho_L(T)$, depending then on its rank instead of its dimensionality. This is a desirable feature for our Trotter-based setup: while $\rho_L(T)$ is a $2^{T/\delta t}\times 2^{T/\delta t}$ matrix, its rank is expected to remain finite and depend on $T$ in the limit $\delta t\rightarrow 0$.

\cref{fig:Delta_nm} displays the evolution of the relative Rényi scalings for several values $6 \geq n> m \geq 2$,  both in the integrable and ergodic regimes. The expected linear growth is obtained in the integrable case.  
This suggests that the spectrum of $\rho_L(T)$ is predominantly flat, with a rank that grows exponentially in time.
Once integrability is broken, we observe a non-monotonic behavior for all R\'enyi scaling. A maximum is attained, qualitatively located at the beginning of the mesoscopic transient period identified in the main text. The position of the maximum shifts slightly towards earlier times as $m$ increases, and to a lesser extent as $n$ does. The ($h=0.5$, $g=0.9$) example in \cref{fig:Delta_nm} has the slowest evolution. In this case, we are only able to establish that the R\'enyi scalings reach a plateau, compatible with a possible subsequent decay that, however, lies outside our accessible time window. 

A very interesting feature can be observed in the fastest evolving example ($h=1$, $g=0.9$). After the maximum, the R\'enyi scalings with $m>2$ reach a minimum and start growing again. Our numerics do not allow us to discern whether they attain a plateau or a second maximum. This effect is more pronounced the larger $n$ and $m$. For $m=2$ it is nevertheless absent in the simulated time window. 
The scalings $\Delta_{2n,2}$ equals $(n-1)$ times the Rényi$-n$ entropy of  the pure state $\bra{L}/\sqrt{\langle{L}\ket{L}}$ with respect to the forward--backward bipartition, and hence measure entanglement between contours. Their monotonic decay  after the maximum indicates that the entanglement between contours decreases along the transient period. 

\section{Toy model for the influence functional spectrum}
\label{SM:toy}

We present now a toy model for the spectrum of $\rho_L$, able to qualitatively reproduce the main features of the R\'enyi scalings reported in \cref{fig:Delta_nm}. In the simplest version of the toy model, we assume that the spectrum consists of a distinct first eigenvalue plus $N$ additional ones with equal value  
\begin{equation}
    \lambda_1=e^{-\gamma T^\alpha} \ , \hspace{1cm} \lambda_N =\frac{1-\lambda_1}{N} \ ,
    \label{eigenmodel}
\end{equation}
with $0<\alpha<1$ and $N$ increasing exponentially with time, such that $\lambda_1$ has a slower long time decay. The value $\alpha=1/2$ corresponds to a diffusive behavior, with $\gamma$ playing the role of diffusive constant.
The Rényi entropies $S_n$ with $n\geq 2$ are dominated by the contribution of the largest eigenvalue, and thus grow asymptotically as $T^\alpha$. The entanglement entropy is instead governed by the exponentially many equal eigenvalues and increases linearly with time.

In this toy model, the Rényi scalings \eqref{deltanm} are given by
\begin{equation}
   \Delta_{n,m}=
   -\log\left( 1+ \frac{\big(1/\lambda_1 -1\big)^n}{ N^{n-1}}\right) +\frac{n}{m} \log\left( 1+ \frac{\big(1/\lambda_1 -1\big)^m} {N^{m-1}}\right)\ .
\end{equation}
This expression attains a maximum when $\lambda_1$ starts dominating over the collective contribution of the $N$ equal eigenvalues. From then on, it decreases monotonically to zero, reflecting that the spectrum becomes effectively more and more sparse with time. Increasing both $n$ and $m$ causes the maximum to occur earlier and accelerates the subsequent decay.

\begin{figure}
\centering
\includegraphics[width=1.0\linewidth]{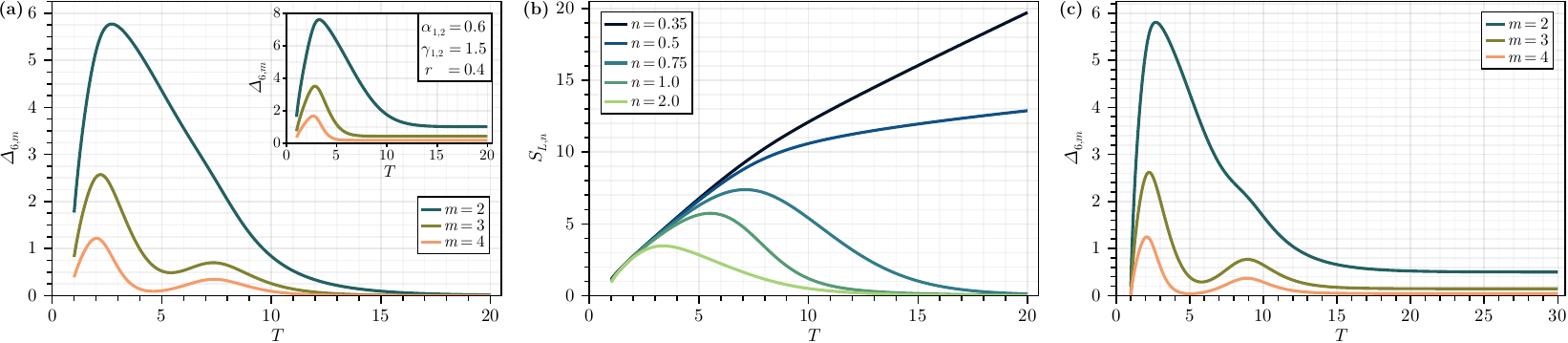}
\caption{Dynamics of the toy model, for the parameter set $\alpha_1=0.6$, $\gamma_1=1.5$, $\alpha_2=0.4$, $\gamma_1=1.25$, $r=0.1$ and $N=2^{2T}-2$. \textbf{(a)} Evolution of the Rényi scaling $\Delta_{6,m}$ for $m\in\{2,\,3,\,4\}$. Inset: $\Delta_{6,m}$ for the different set of parameters $\alpha_{1,2}=0.6$, $\gamma_{1,2}=1.5$ and $r=0.4$. \textbf{(b)} R\'enyi entropies $S_{L,n}$ of
$\frac{|L\rangle \langle L|}{\langle L|L\rangle}$ in the forward versus backward bipartition for several values of $n$. \textbf{(c)} Rényi scaling $\Delta_{6,m}$ including a third distinct eigenvalue $\lambda_3$ according to \cref{eigenmodel3}, with $r'=0.3$.}
\label{fig:toy}
\end{figure}

The monotonic decay to zero after the maximum does not agree however with \cref{fig:Delta_nm}, showing the limitation of the simple toy model \eqref{eigenmodel}. A richer evolution pattern can be obtained by allowing instead for two slowly decaying eigenvalues
\begin{equation}
    \lambda_1= (1-r) e^{-\gamma_1 T^{\alpha_1}} \ , \hspace{0.7cm} \lambda_2=r \, e^{-\gamma_2 T^{\alpha_2}} \ , \hspace{0.7cm}  \lambda_N =\frac{1-\lambda_1-\lambda_2}{N} \ ,
    \label{eigenmodel2}
\end{equation}
with $0<\alpha_1, \alpha_2 <1$, and $0<r<1$ controlling the relative weights of the first two eigenvalues at $T=0$. Let us consider $\alpha_1>\alpha_2$, or $\alpha_1=\alpha_2$ and $\gamma_1>\gamma_2$, such that $\lambda_2$ has the slowest asymptotic decay. We also assume $r<1/2$ such that $\lambda_1$ is larger than $\lambda_2$ at short times. Choosing the parameters appropriately along these lines, $\Delta_{n,m}$ attains a first maximum when $\lambda_1$ dominates over the joint contribution of $\lambda_2$ plus the $N$ equal eigenvalues, after which it decreases. A minimum is then reached when the contributions of $\lambda_1$ and $\lambda_2$ start having a similar weight. A second maximum occurs when $\lambda_2$ surpasses the contribution of $\lambda_1$, which finally gives way to a monotonic decay to zero, see \cref{fig:toy}(a).
The second maximum turns out to be more pronounced the larger $m$ is. For $m=2$ and a generic choice of parameters along the previous lines, the second maximum is either weak or disappears. These features capture qualitatively those of the Ising model simulations in \cref{fig:Delta_nm}. 
This suggests that the long mesoscopic transient period is associated with a slow reorganization of the influence functional spectrum.

The scaling $\Delta_{2n,2}$ equals $n-1$ times $S_{L,n}$, the Rényi-n entropy of the forward versus backward bipartition associated to the pure state $|L\rangle$, the vectorized form of $\rho_L$. This is immediately seen by noticing that the eigenvalues of the reduced density matrix obtained by tracing $\frac{|L\rangle \langle L|}{\langle L|L\rangle}$ over the backward temporal entries
are ${\lambda_i^2 / \tr \rho_L^2}$, with $\lambda_i$ the eigenvalues of $\rho_L$. The toy model shows clearly how this spectrum amplifies the sparsity of $\rho_L$, inducing the non-monotonic behavior of \eqref{deltanm}.  

The decay to zero of the higher R\'enyi entropies of the pure state with respect to the forward versus backward bipartition indicates that the entropy between contours decays in time. We could wonder about the efficiency of folding algorithms in systems with this behavior.  In order to address the representability of a state as MPS, however, also R\'enyi entropies with $n<1$ have to be considered~\cite{Schuch_2008}.  Let us return to the toy model, which leads to the following long-time behavior when $n<1$
\begin{equation}
   S_{L,n} \to \frac{1}{1-n} \log \left( 1+ \frac{1}{(\lambda_1^{2n}+\lambda_2^{2n}) \, N^{2n-1}} \right) \ .
    \label{renyiL}
\end{equation}
We observe that $S_{L,n}$ also vanishes asymptotically when $n>1/2$. Notice that the same property holds for the associated entanglement entropy due to the monotonicity in $n$ of the R\'enyi entropies.
When $n=1/2$, and with the above assumptions on the parameters of the toy model, \eqref{renyiL} grows as $T^{\alpha_2}$ at long times, while it does so linearly for $n<1/2$, see \cref{fig:toy}(b). 
Hence, the non-monotonic behavior of the Rényi scalings found in our simulations 
does not imply an automatic limitation for folding algorithms, its potential implications deserving further study.

Finally, notice that the time window accessible to our numerics does not allow discerning whether after the second maximum there is a plateau or a decay to zero of the R\'enyi scaling, see \cref{fig:toy}(a). The decay to zero present in the toy models can be avoided if at least two eigenvalues with the same slow decay dominate at long times. For example, we could add to \eqref{eigenmodel2} a third distinct eigenvalue such that
\begin{equation}
    \lambda_2=r (1-r') \, e^{-\gamma_2 T^{\alpha_2}} \ , \hspace{1cm} \lambda_3=r r' \, e^{-\gamma_2 T^{\alpha_2}} \ ,
    \label{eigenmodel3}
\end{equation}
adjusting accordingly $\lambda_N$. With this simple modification, the R\'enyi scalings would tend asymptotically to
\begin{equation}
    \Delta_{n,m} \to -\log \frac{(1-r')^n +r'^n}{((1-r')^m +r'^m)^{\frac{n}{m}}} \ .
\end{equation}

\section{Diffusive behavior in ergodic systems with conserved quantities}
\label{SM:Diff_non_random}

The objective of this section is twofold. First, we investigate whether the truncation of matrix product states and the Trotterization of the time-evolution operator break the conditions required to observe diffusive scaling at long times in quantities that probe spatial partitions. Second, we study the behavior of the amplitude \(A(t)\), introduced in reference~\cite{zhou2020diffusive}, for some of the cases considered in this paper, to assess the time scales associated with the onset of the diffusive regime in the systems analyzed here.

In Ref.~\cite{zhou2020diffusive}, the authors characterize the amplitude
\begin{equation}
    A(t) = \operatorname{tr}\left( U(t)\rho_0 U_L^\dagger(t)U_R^\dagger(t) \right)
    \label{eq:evo_ampl}
\end{equation}
for a one-dimensional system of \(N\) spins. Here, \(U(t)\) denotes the time evolution of the full system, while \(U_L(t)\) and \(U_R(t)\) denote the independent time evolution of the left and right halves of the system, respectively. In other words, if the full dynamics is generated by \(U(t)=\exp(-iHt)\), then \(U_L(t)=\exp(-iH_Lt)\) and \(U_R(t)=\exp(-iH_Rt)\), where \(H_L\) and \(H_R\) are the restrictions of the Hamiltonian to the left and right halves of the chain.

If the dynamics has a \(U(1)\) symmetry generated by an extensive conserved charge
\[
S=\sum_i s_i,
\]
with local density \(s_i\), such that \(\left[e^{i\alpha S},U(t)\right]=0\) for all \(\alpha\), then the amplitude \(A(t)\) is expected to decay diffusively at long times,
\begin{equation}
    A(t)\propto e^{-\sqrt{D t}} .
\end{equation}
By contrast, in systems without such a conserved quantity, the decay is expected to be ballistic,
\begin{equation}
    A(t)\propto e^{-D' t}.
\end{equation}
The authors in Ref.~\cite{zhou2020diffusive} argue that ergodic Hamiltonian dynamics should fall into the diffusive class, since energy conservation plays the role of the conserved quantity.
 
We follow the setup of Ref.~\cite{zhou2020diffusive}. We consider the Ising model for a chain of \(L=22\) qubits with Hamiltonian
\begin{equation}
    H=-J\sum_i^L X_iX_{i+1}-\sum_i^L\left(h Z_i+g X_i\right) - JX_1-JX_L .
\end{equation}
The last two terms fix the boundary conditions and reduce finite-size effects.

\begin{figure}[h]
\centering
\includegraphics[width=\textwidth]{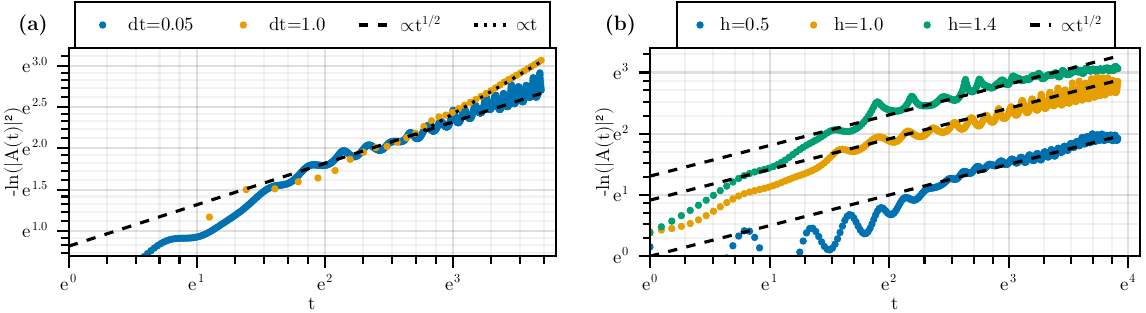}
\caption{Log-Log plots of the logarithm of the squared evolution amplitude defined in \cref{eq:evo_ampl}. Panel \textbf{(a)} corresponds to the same setup as in Ref.~\cite{zhou2020diffusive}, with \(L=22\), \(h=0.809\), and \(g=-0.9045\). The dashed line corresponds to the fit \(\ln{|A(t)|^2} = -2.26t^{0.502}\), while the dotted line corresponds to a linear fit, expected in Floquet dynamics. Panel \textbf{(b)} corresponds to some of the cases studied in this paper, with \(g=0.9\) and different transverse fields. In the same panel, we overlay different power-law behaviors. 
}
\label{fig:zhou}
\end{figure}

In \cref{fig:zhou}(a), we see that Trotterization and matrix-product-state truncation do not modify the expected scaling of \(A(t)\). In blue and orange, we show the time evolution of \(\ln |A(t)|^2\) for the Ising model with \(h=0.809\) and \(g=-0.9045\), a parameter regime where the model is expected to be strongly ergodic~\cite{kim2013}. For Hamiltonian evolution with \(dt=0.05\) (blue), we find
\[
\ln |A(t)|^2 \approx -2.26 t^{0.502},
\]
in agreement with the diffusive prediction of Ref.~\cite{zhou2020diffusive}. By contrast, for Floquet evolution with \(dt=1\) (orange), \(\ln |A(t)|^2\) displays ballistic behavior at late times.

In \cref{fig:zhou}(b), we show \(\ln |A(t)|^2\) for a set of Ising-model parameters considered in the main text. We focus on \(g=0.9\) and \(h=0.5,1,1.4\) (blue, orange, and green circles, respectively). On top of the numerical data, we overlay black dashed lines corresponding to the power-law behavior expected in the presence of a conserved quantity,
\[
\ln |A(t)|^2 \propto \sqrt{t}.
\]
For \(h=0.5\) and \(h=1\), the behavior of \(\ln |A(t)|^2\) is compatible with the \(\sqrt{t}\) scaling. However, for \(h=1.4\), this is no longer the case. The results are robust for different system sizes (we have checked $L=22,26,30$ and the relative differences of \(\ln |A(t)|^2\) are less than 1\%). Furthermore, we have checked that the results are sufficiently converged with bond dimension, and we find the same behavior for $\chi=200,300$, and $500$.

These results reinforce the conclusions of the main text: non-integrable Hamiltonian dynamics can exhibit significant deviations from the random-circuit paradigm, even when the latter is enriched by conserved quantities.

\ifdefined\IsMainDocument
\else
    \bibliography{References.bib}
    \end{document}
\fi
\fi

\end{document}